# Comment on "Observation of the Wigner-Huntington transition to metallic hydrogen"

A recent paper of Dias and Silvera (DS) [1] reports on production of metallic hydrogen in a diamond anvil cell at 495 GPa at 5.5 and 83 K. The results are implied to have a great impact on "energy and rocketry". Here we argue that the presented (very scarce) results are contradictory with the presented experimental description making their claims unsupported experimentally. Moreover, the proposed implications are highly speculative making this paper very confusing for a broad audience [2]. Elucidating the claims and the related implications is important for building a coherent picture that is currently emerging as the results of theoretical calculations at various levels and experimental investigations employing static and dynamic compression techniques (*e.g.* Refs. [*3-7*]). There is no doubt that hydrogen metallizes at high pressures. But this does not make all claims about reaching this state immediately valid. Scientific community would like to learn at what conditions hydrogen metallizes, what is the nature of the conducting state [8] and its properties (*e.g.* superconductivity). Here we argue that the presented data do not provide any reliable information on this.

There is a lot of interest in the high pressure community to the experimental evidence of metallic hydrogen phase claimed in the publication at hand. We briefly reiterate here that the pressure in the experiment of DS is not measured using the accepted by the community standards: there is no continuity in the IR- Raman data presented, and the Raman data at the claimed 495 GPa are of low quality and do not look unambiguously as a Raman signal from the stressed anvil as it is reported in other works [*5, 7*], so we express a doubt that DS were even in a close vicinity of the claimed pressure. Previously used extrapolations of the pressure-load curves cannot serve as a reliable tool any more: moreover, the linear curves of this kind are totally unrealistic! (*e.g.* Ref. [9])

DS report visual observations of a very shiny material in the DAC cavity. Without specifying what could cause such observations we note that there is no diagnostics of hydrogen presence in the cavity at these conditions, for example, using continuity arguments based on the monitoring of the physical properties of hydrogen (*e.g.* Ref. [*5, 7*]).

DS also report reflectivity measurements of their "Wigner-Huntington" state of hydrogen. They use the diamond absorption correction of Ref. [*10*], to offset the unknown absorption of their diamonds at high pressure (Fig. 1). However, the correction has been applied erroneously as according to Vohra [*10*] at 405 nm (3.06 eV, see also Fig. S4) the optical density of the diamond anvil should be around 6 (extrapolated to 495 GPa) leaving almost no light to pass. This contradicts DS observations of about 50% of raw reflectivity at 405 nm suggesting that their signal is spurious (not from the sample) and/or they have not reached the claimed pressure.

In summary, we refute the claim of DS that "We have produced atomic metallic hydrogen in the laboratory at high pressure and low temperature". They rather reported artifact of their measurements at the unknown (likely at much lower than claimed) pressures and their observations have nothing to do with the properties of metallic hydrogen, which will be a topic for the research to come.

Alexander F. Goncharov and Viktor V. Struzhkin
Geophysical Laboratory, Carnegie Institution of Washington, Washington DC, 20015

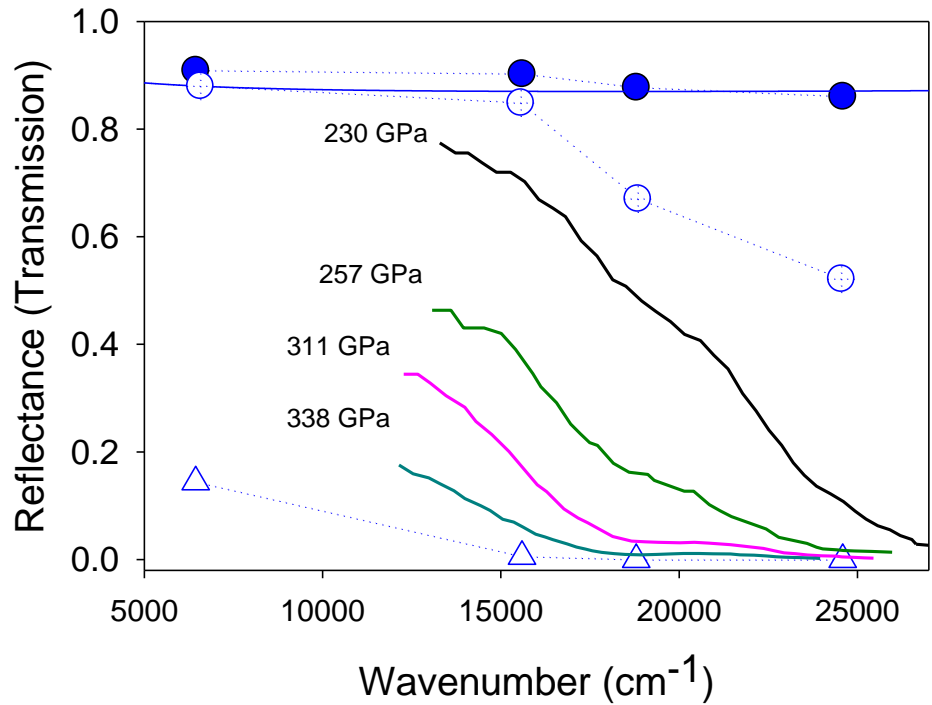

**Fig. 1. Reflectance of hydrogen at 495 GPa determined by DS (filled circles- corrected, open circles – uncorrected) in comparison with the diamond transmittance from Vohra [*10*] – at 230, 257, 311, and 338 GPa (broadly consistent with the results of Goncharov *et al.* [*11*]) used to correct the data for the diamond absorption at high pressures. The proper correction (estimated from Vohra's data extrapolated to 495 GPa) suggests that uncorrected results should be much lower than reported by DS– these are shown as open triangles. These estimates strongly suggest that the reflected signal presented by DS was spurious and/or pressures were much lower than claimed, below 230 GPa.**

**Literature**
[1] R.P. Dias, I.F. Silvera, Observation of the Wigner-Huntington transition to metallic hydrogen, Science, DOI 10.1126/science.aal1579(2017).
[2] V.V. Brazhkin, A.G. Lyapin, Metastable high-pressure phases of low-Z compounds: creation of a new chemistry or a prompt for old principles?, Nat Mater, 3 (2004) 497-500.
[3] M.D. Knudson, M.P. Desjarlais, A. Becker, R.W. Lemke, K.R. Cochrane, M.E. Savage, D.E. Bliss, T.R. Mattsson, R. Redmer, Direct observation of an abrupt insulator-to-metal transition in dense liquid deuterium, Science, 348 (2015) 1455-1460.
[4] J.M. McMahon, M.A. Morales, C. Pierleoni, D.M. Ceperley, The properties of hydrogen and helium under extreme conditions, Reviews of Modern Physics, 84 (2012) 1607-1653.
[5] M.I. Eremets, I.A. Troyan, A.P. Drozdov, Low temperature phase diagram of hydrogen at pressures up to 380 GPa. A possible metallic phase at 360 GPa and 200 K, arXiv:1601.04479 [cond-mat.mtrl-sci], DOI (2016).
[6] J. Chen, X.-Z. Li, Q. Zhang, M.I.J. Probert, C.J. Pickard, R.J. Needs, A. Michaelides, E. Wang, Quantum simulation of low-temperature metallic liquid hydrogen, Nature Communications, 4 (2013) 2064.

[7] P. Dalladay-Simpson, R.T. Howie, E. Gregoryanz, Evidence for a new phase of dense hydrogen above 325 gigapascals, Nature, 529 (2016) 63-67.
[8] E. Babaev, A. Sudbo, N.W. Ashcroft, A superconductor to superfluid phase transition in liquid metallic hydrogen, Nature, 431 (2004) 666-668.
[9] R.J. Hemley, H.-k. Mao, G. Shen, J. Badro, P. Gillet, M. Hanfland, D. Häusermann, X-ray Imaging of Stress and Strain of Diamond, Iron, and Tungsten at Megabar Pressures, Science, 276 (1997) 1242-1245.
[10] Y.K. Vohra, Isotopically pure diamond anvil for ultrahigh pressure research, in Proceedings of the XIII AIRAPT International Conference on High Pressure Science and Technology (1991, Bangalore, India, 1991). (1991).
[11] A.F. Goncharov, E. Gregoryanz, R.J. Hemley, H.-k. Mao, Spectroscopic studies of the vibrational and electronic properties of solid hydrogen to 285 GPa, Proceedings of the National Academy of Sciences, 98 (2001) 14234-14237.